\documentclass{article}

\usepackage{arxiv}

\usepackage[utf8]{inputenc} 
\usepackage[T1]{fontenc}    
\usepackage{hyperref}       
\usepackage{booktabs}       
\usepackage{amsfonts}       
\usepackage{nicefrac}       
\usepackage{microtype}      
\usepackage{lipsum}		
\usepackage{graphicx}
\usepackage{doi}

\usepackage[
    backend=biber,
    style=numeric-comp,
    natbib=true,
    url=false,
    doi=true,
    eprint=false
]{biblatex}
\usepackage{hyperref}
\hypersetup{
    colorlinks,
    allcolors=black,
    pdftitle={Privacy-Preserving Linkage of Distributed Datasets using the Personal Health Train},
    pdfsubject={cs.DC},
    pdfauthor={Maximilian Jugl, Sascha Welten, Yongli Mou, Yeliz U. Yediel, Oya D. Beyan, Ulrich Sax, Toralf Kirsten},
    pdfkeywords={Distributed Computing, Distributed Datasets, Incremental Learning, Federated Learning, Record Linkage, Cybersecurity, Data Privacy, Data Protection, Containerization},
}

\usepackage{siunitx}
\usepackage{csquotes}

\title{Privacy-Preserving Linkage of Distributed Datasets using the Personal Health Train}

\date{September 12, 2023}	

\author{ \href{https://orcid.org/0009-0000-8479-1716}{\includegraphics[scale=0.06]{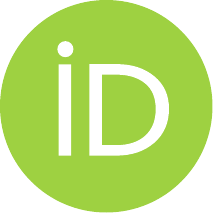}\hspace{1mm}Maximilian Jugl} \\
    Department of Medical Data Science \\
    University Medical Center Leipzig \\
    Leipzig, Germany \\
    \texttt{maximilian.jugl@medizin.uni-leipzig.de} \\
    \And
    \href{https://orcid.org/0000-0001-5570-9672}{\includegraphics[scale=0.06]{orcid.pdf}\hspace{1mm}Sascha Welten} \\
    Chair of Computer Science 5 \\
    RWTH Aachen University \\
    Aachen, Germany \\
    \texttt{welten@dbis.rwth-aachen.de} \\
    \And
    \href{https://orcid.org/0000-0002-2064-0107}{\includegraphics[scale=0.06]{orcid.pdf}\hspace{1mm}Yongli Mou} \\
    Chair of Computer Science 5 \\
    RWTH Aachen University \\
    Aachen, Germany \\
    \texttt{mou@dbis.rwth-aachen.de} \\
    \And
    \href{https://orcid.org/0000-0002-6845-7774}{\includegraphics[scale=0.06]{orcid.pdf}\hspace{1mm}Yeliz U.~Yediel} \\
    Department of Data Science and Artificial Intelligence \\
    Fraunhofer FIT \\
    Sankt Augustin, Germany \\
    \texttt{yeliz.ucer.yediel@fit.fraunhofer.de} \\
    \And
    \href{https://orcid.org/0000-0001-7611-3501}{\includegraphics[scale=0.06]{orcid.pdf}\hspace{1mm}Oya D.~Beyan} \\
    Institute for Medical Informatics \\
    Faculty of Medicine and University Hospital Cologne \\
    Cologne, Germany \\
    \texttt{oya.beyan@uni-koeln.de} \\
    \And
    \href{https://orcid.org/0000-0002-8188-3495}{\includegraphics[scale=0.06]{orcid.pdf}\hspace{1mm}Ulrich Sax} \\
    Department of Medical Informatics \\
    University Medical Center Göttingen \\
    Göttingen, Germany \\
    \texttt{ulrich.sax@med.uni-goettingen.de} \\
    \And
    \href{https://orcid.org/0000-0001-7117-4268}{\includegraphics[scale=0.06]{orcid.pdf}\hspace{1mm}Toralf Kirsten} \\
    Department of Medical Data Science \\
    University Medical Center Leipzig \\
    Leipzig, Germany \\
    \texttt{toralf.kirsten@medizin.uni-leipzig.de} \\
}

\renewcommand{\headeright}{Preprint}
\renewcommand{\shorttitle}{Privacy-Preserving Linkage of Distributed Datasets using the Personal Health Train}

\begin{document}
\maketitle

\begin{abstract}
With the generation of personal and medical data at several locations, medical data science faces unique challenges when working on distributed datasets.
Growing data protection requirements in recent years drastically limit the use of personally identifiable information.
Distributed data analysis aims to provide solutions for securely working on highly sensitive data while minimizing the risk of information leaks, which would not be possible to the same degree in a centralized approach.
A novel concept in this field is the Personal Health Train (PHT), which encapsulates the idea of bringing the analysis to the data, not vice versa.
Data sources are represented as train stations.
Trains containing analysis tasks move between stations and aggregate results.
Train executions are coordinated by a central station which data analysts can interact with.
Data remains at their respective stations and analysis results are only stored inside the train, providing a safe and secure environment for distributed data analysis.

Duplicate records across multiple locations can skew results in a distributed data analysis.
On the other hand, merging information from several datasets referring to the same real-world entities may improve data completeness and therefore data quality.
In this paper, we present an approach for record linkage on distributed datasets using the Personal Health Train.
We verify this approach and evaluate its effectiveness by applying it to two datasets based on real-world data and outline its possible applications in the context of distributed data analysis tasks.

The source code for the services and analysis scripts mentioned in this paper is open source and available from the authors\footnote{The source code is located at \url{https://gitlab.com/ul-mds/record-linkage/infrastructure}.}.
\end{abstract}


\section{Introduction}

\subsection{Distributed datasets}

In the medical domain, new data is generated on a daily basis.
We observe various types of data, such as structured data in the shape of electronic health records, textual data from written diagnostic reports, image data from MRI and CRT scans and genetics data for research purposes.
With large amounts of multi-faceted data, multiple perspectives have to be considered at once, such as the challenge of efficiently and effectively managing data, capturing patient data in an appropriate and timely manner, and the eventual analysis of captured data by researchers to support practicing medical personnel in their daily tasks.

In a data analysis undertaking spanning multiple institutions, the common approach is to collect all required data in a central location first.
This is not without its own challenges.
First, with growing data protection requirements, as enforced by the GDPR in the European Union for example~\cite{voigt_eu_2017}, sensitive data has to be carefully anonymized so that re-identification of natural persons is infeasible.
Second, datasets must be submitted over a secure channel, where the dataset size and the communication overhead can greatly impact the time needed for a successful transfer.
Third, with a transmission of data from one domain to another comes a loss of sovereignty for the original data holders, effectively diminishing their say in how the data they provide can and should be used.

As such, a recent paradigm shift in the field of medical data science is to favor distributed data analysis approaches, where data analysis is performed at the respective data sources.
Over the last decade, research in this area gave way to novel concepts and implementations to enable data analysis while respecting data ownership.
Galaxy~\cite{goecks_galaxy_2010} enables the effective communication of analysis tasks between multiple parties.
Local installation of their software allows data providers to work on their data without exposing it to others.
ELIXIR~\cite{crosswell_elixir_2012} acts as an all-in-one provider of data analysis solutions, represented as several platforms, for scientific purposes in Europe.
A central hub near Cambridge handles administrative tasks and coordinates nodes in \num{23} European countries, where a node is accessed by multiple institutions within that country.
DataSHIELD~\cite{gaye_datashield_2014} takes this one step further by allowing researchers to write their analysis tasks as R scripts that are executed in a distributed way using a custom client-server infrastructure.
PADME~\cite{welten_privacy-preserving_2022}, PHT-meDIC~\cite{herr_bringing_2022} and VANTAGE6~\cite{moncada-torres_vantage6_2021} are implementations of the Personal Health Train~\cite{beyan_distributed_2020}, which frees data analysts from constraints on how to define their data analysis tasks.

\subsection{Personal Health Train}

\begin{figure}[!ht]
	\centering
    \includegraphics[width=.6\textwidth]{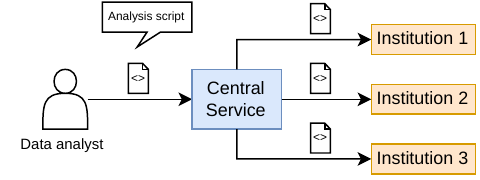}
	\caption{Overview of the PHT using the PADME implementation as an example. A data scientist submits an analysis script to the Central Service, which packages it into a train using container-based technologies and orchestrates the execution between institutions, called stations. Results are aggregated and reported back to the data scientist.}
	\label{fig:pht-overview}
\end{figure}

The Personal Health Train (PHT) follows the analogy of a train which moves between stations, where the train represents a data analysis task and the stations represent the respective data holding organizations~\cite{beyan_distributed_2020}.
All current PHT implementations use a container-based approach to package analysis tasks and to transmit them between institutions, which allows data scientists to freely choose the tools they need to perform their analysis tasks.
A broad overview of the PHT workflow is shown in Fig~\ref{fig:pht-overview} using the PADME implementation as an example.

A data scientist interacts with the Central Service (CS).
They submit their analysis task as an executable script to the CS, where it is then packaged into a train class with all required dependencies.
The train class acts as a template for trains that are created from it.
Once the train class has been created, the data scientist can select it and a number of target stations.
A train, which is effectively an executable container, is then created from the selected train class and the participating stations are informed of the train.
The station order is defined by the data scientist.

Station administrators run their stations through a piece of on-premise software.
They may pull trains or reject them if they suspect them to be malicious.
Before a train is executed, the station administrators have the option to supply variables for the execution of the analysis script.
This is necessary to provide URLs to services that provide data, as well as authentication credentials for access to these services and other configuration parameters.
Once train execution has finished, the results are packaged into the train and sent off to the next station.

After the train has been executed at all stations, the final train is sent back to the CS.
Results are extracted from the train and provided to the data scientist.
At no point did the data scientist ever obtain access to confidential data.
Access to sensitive data sources remains at the participating institutions.
The only PHT component that the data scientist has to interact with to perform a distributed data analysis task is the CS.

Multiple recent use cases using the PHT prove its potential as a functional mean to enabling distributed data analysis~\cite{mou_distributed_2021, welten_synthetic_2022}.
In cases where a centralized study has been performed in the past, running the same study in a decentralized way using the PHT shows no to minimal deviation in the quality of the obtained results~\cite{welten_multi-institutional_2022}.

\subsection{Problem statement}

In distributed datasets, it is possible for duplicates to manifest that refer to the same real-world entity.
Within the medical domain, these entities are usually patients that seek treatment at multiple locations.
In a patient's lifetime medical history, it is common to seek treatment at multiple healthcare providers, such as being referred by one's own general practitioner to obtain a verified diagnosis that they cannot make, or seeking special treatment for a rare disease type.

The process of identifying similar records within one or multiple datasets is called data matching or record linkage~\cite{christen_data_2012}.
Duplicates in distributed datasets can skew the outcome of an analysis task, yet they may also present the opportunity of fusing multiple information sources together.
Though record linkage algorithms have been extensively studied and applied to centralized datasets, very few have applied them in a distributed environment.
This is rooted in the inherent challenges of record linkage.

Linkage using a permanent identifier, like a person's health insurance number, passport ID, identity card number or social security number would facilitate this process tremendously, yet they are subject to strict data privacy and data sharing protections.
Additionally, healthcare providers may only store a subset of the aforementioned identifiers, shrinking the potential overlap in distributed datasets.

As such, record linkage relies on combining a multitude of so-called quasi-identifiers (QIDs) to identify similar records~\cite{winkler_chapter_2009}.
These are stable identifiers that are unlikely to change over a long period of time, such as a patient's first and last name, sex and birth date.
Yet since combining these identifiers allows one to uniquely identify a person with high precision, they are subject to the same data protection guidelines as the mentioned identifiers.
Releasing identifying information past institutional borders for the purpose of record linkage therefore poses a great risk for data providers and the privacy of the entities that are supposed to be linked.

Using the distributed data analysis execution platform that the PHT provides, we propose an approach to record linkage with distributed datasets which does not require data providers to share identifying data outside their own borders.

\subsection{Related work}

Though all of the following works present ways on how to integrate record linkage into real-life environments, just a select few propose workflows on distributed datasets.
None so far have sought the PHT as their platform of choice.

Randall et al.~\cite{randall_privacy-preserving_2014} performed an evaluation of a Bloom filter based record linkage protocol on datasets with a total of \num{26} million hospital admission records.
They showed that Bloom filters can scale well to large datasets and provided a workflow to improve scalability without severely affecting linkage quality.
However, the setup was performed locally and not in a distributed environment.

Yigzaw et al.~\cite{yigzaw_secure_2017} designed a record linkage protocol based on Bloom filters for disease surveillance across three laboratories that specialize in the testing for strains of influenza.
Their approach manages to stay performant while ensuring security in a setting with semi-honest adversaries.

Nguyen et al.~\cite{nguyen_privacy-preserving_2020} performed an evaluation of the GRHANITE software in a public health surveillance system.
They used two separate gold standard datasets sourced from several clinical sites in Australia, including uniquely identifiable attributes such as electronic medical record numbers.
The authors opted for a deterministic linkage approach which, while reliable in most cases, provides limited error tolerance due to the nature of this approach.

Stammler et al.~\cite{stammler_mainzelliste_2020} extended the medical record database software Mainzelliste with a modified version of the EpiLink software by Contiero et al.~\cite{contiero_epilink_2005} to enable secure multi-party privacy-preserving record linkage.
Their linkage algorithm is based on Bloom filters, but the integrity and privacy of shared data between participating clients is preserved using a tunneled connection.
However, the integration into Mainzelliste imposes a strong software dependency that may not be implementable in certain environments.

Nóbrega et al.~\cite{nobrega_blockchain-based_2021} recently published a novel protocol using Bloom filters using Blockchain technology and applied it to real-world datasets.
Their main motivation for this approach was to provide security beyond the commonly considered honest-but-curious adversary model.
Since then, Christen et al.~\cite{christen_critique_2022} published an attack on the protocol which has been considered and mitigated by the original authors~\cite{nobrega_explanation_2022}.

\section{Materials and methods}
\label{sec:method}

\subsection{Overall approach}
\label{sec:method-approach}

\begin{figure}[!ht]
	\centering
    \includegraphics[width=\textwidth]{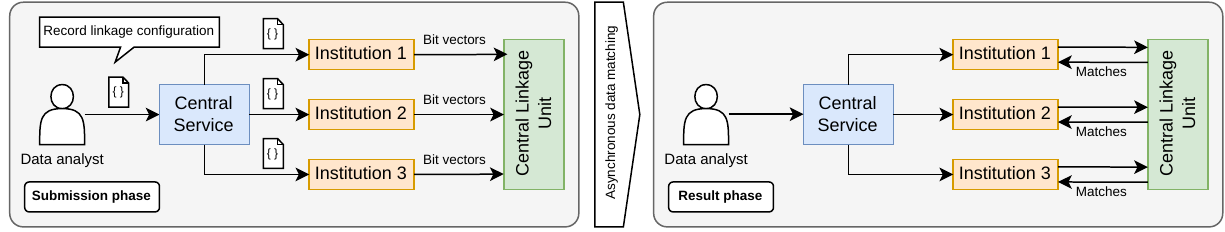}
	\caption{Overview of the record linkage execution within the PHT. In the first phase, stations are prompted to encode records into a non-reversible bit vector form and to submit these to a Central Linkage Unit. After matching has concluded, results are fetched from the Central Linkage Unit at each station.}
	\label{fig:pprl-phases}
\end{figure}

We developed an approach for performing record linkage on distributed datasets using the PHT.
Fig~\ref{fig:pprl-phases} demonstrates the two-phase record linkage execution within the PHT.

In the \enquote{submission} phase, the data analyst sends their record linkage script along with a configuration file to the central service.
They select the participating stations and dispatch the record linkage train.
At each station, the train starts the process of sourcing relevant records, pre-processing and masking them into a non-reversible bit vector form while preserving similarities between near-identical records.
These bit vectors are then sent to a Central Linkage Unit (CLU), which asynchronously performs matching on the provided bit vectors.

In the \enquote{result} phase, the data analyst dispatches the same train for a second time.
Once the train arrives at the station, it prompts the retrieval of results from the CLU for that particular station.
The train itself receives identified matches in a pseudonymized form, so that the train never comes into contact with any personal data.
These results can be used to perform informed decisions on how to treat duplicates in a subsequent data analysis task.
The following sections describe the process in more detail.

\subsection{Train execution}
\label{sec:method-train}

\begin{figure}[!ht]
	\centering
    \includegraphics[width=.8\textwidth]{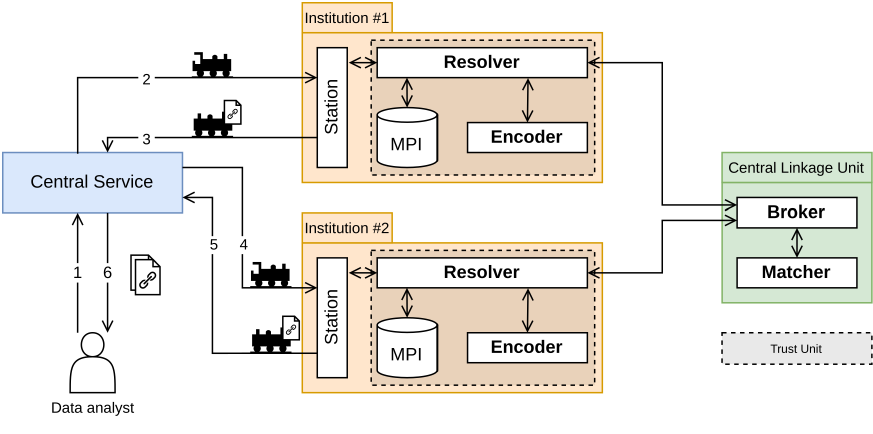}
	\caption{Integration of record linkage services into the PHT infrastructure. Custom components are highlighted in bold. Trains carry the scripts necessary to communicate with the Resolver service at each station and report identified matches back to the Central Service.}
	\label{fig:pprl-infra}
\end{figure}

We developed several standalone web services that enable record linkage with distributed datasets: a Resolver, Encoder, Broker and Matcher service.
An architectural overview of their integration into the PADME PHT implementation is presented in Fig~\ref{fig:pprl-infra}.
Although our research focuses on the interoperability with the PHT, our services do not rely on any PHT component and can be freely made to fit with any architecture where client-server interactions are possible.

\paragraph{Components}

Our services are split up into station-side and central components.
The Resolver service is hosted on-premise at every participating station.
Its purpose is to look up study-specific pseudonyms in the station's Master Patient Index (MPI).
For our purposes, we opted for E-PIX as the patient database and gPAS as the pseudonymization service from the MOSAIC suite of tools to act as an MPI in combination~\cite{bialke_mosaic_2015, bialke_workflow-driven_2015}.
These services offer an intuitive user interface for data providers, as well as a SOAP interface for integration into other software.
However, the Resolver offers flexibility to integrate with other MPI software solutions.
The Resolver leverages the Encoder service, which performs data pre-processing and masking based on a hash-based approach using Bloom filters~\cite{schnell_novel_2011} on the records obtained from the MPI.
The result is a bit vector for every record which obfuscates the original data while preserving their similarity.
The CLU is composed of the Broker service which takes in bit vectors from all stations.
Similarities between bit vectors are computed by the Matcher service and reported back to the Broker, which provides results to the participating stations.

\paragraph{Workflow}

As mentioned before, a record linkage execution within the PHT consists of two phases.
These two phases make up a matching session.
In the first \enquote{submission} phase, the data analyst who wants to perform record linkage selects the corresponding train image at the Central Service, the participating stations and provides the necessary configuration which is packaged into the train.
The record linkage configuration consists of a random session identifier and a matching threshold between \SI{0}{\percent} and \SI{100}{\percent}.
Upon arrival of the train, station administrators provide a link to the Resolver service hosted at the station, as well as a list of study-specific pseudonyms.

Once train execution commences at the station, an executable script inside the train submits the supplied pseudonyms as well as the record linkage configuration to the Resolver service at the station.
The Resolver looks up the provided pseudonyms in the MPI and converts the personal data records into bit vectors using the Encoder service.
Finally, the Resolver generates a random client identifier, ties it to the session identifier locally, and passes both with the record linkage configuration and the list of bit vectors to the CLU.
The first phase has concluded once all participating stations have performed this process.

Between phases, the CLU performs asynchronous matching using the bit vectors submitted by the stations.
The submitted station and client identifiers are used by the Broker service to tell bit vectors from different stations apart, as well as consolidating bit vectors that are supposed to be matched against one another.
Given bit vectors from two separate stations, the Matcher performs a cross product on them and computes their similarity using the Jaccard index.
Similarities above the threshold specified in the PPRL configuration are reported back to the Broker, which aggregates these results.

In the second \enquote{result} phase, the data analyst sends the same PPRL train for a second iteration to all participating stations.
The script inside the train contacts the Resolver hosted at the station.
The Resolver looks up the client identifier generated in the previous phase using the session identifier and requests results from the Broker.
Finally, the Resolver passes the results along to the train, converting them back into their pseudonymized form.
This means that by the end of the second phase, every station will have a list of pseudonyms that refer to records that have a match with at least another record at a different station.
The train itself would have had no access to sensible data during the entire session.

\subsection{Data preprocessing}
\label{sec:method-preproc}

The QIDs withing a record are first preprocessed at their respective stations.
The aim is to convert all records into a unified representation across all stations, so that record linkage is successful even in the presence of different character encodings and naming conventions.
This is done by performing the following series of steps for every field of a record:

\begin{enumerate}
\item Ligatures (e.g. ß, æ, œ) are replaced with their non-ligature forms (e.g. ss, ae, oe).
\item Diacritics are separated from the characters they apply to by performing Unicode normalization in the KD form.
\item Non-ASCII characters are removed.
\item All characters are converted to lowercase.
\item Multiple consecutive whitespaces are reduced into one.
\end{enumerate}

\subsection{Masking technique}
\label{sec:method-mask}

Masking is the key step in ensuring that the privacy of the matched records is preserved.
If all records are sent to the CLU as they are, then the CLU administrator has access to all records across all stations, posing a big risk of a potential private data breach.
Even if it is assumed that the CLU is hosted by a trustworthy third party, malicious actors could obtain access to the original records if the CLU is ever compromised.
As such, it is necessary to make sure that the records are transmitted in a masked form such that re-identification attacks on masked records are computationally infeasible.

Bloom filters have seen wide adoption in privacy-preserving record linkage (PPRL) protocols as a data structure that can obfuscate the data that is inserted into it while preserving similarities.
Since the inception of Bloom filters in the field of PPRL, many security recommendations have been proposed to make them more resilient to re-identification attacks~\cite{kuzu_constraint_2011, niedermeyer_cryptanalysis_2014, christen_efficient_2017}.
According to current security recommendations for Bloom filters in PPRL, we chose to implement the following measures.

We used CLKRBF as our masking technique~\cite{vatsalan_evaluation_2014}.
The tokenized values of a record are hashed into the same Bloom filter, but the amount of hashes is decided on weights assigned to a record's attributes.
Consequentially, attributes with a higher weight likely occupy more bits in the resulting Bloom filter.
The reason for this approach is that some attributes, like first and last name, have a higher discriminatory power than others, like gender~\cite{durham_composite_2014}.
We computed the weights by generating a random list of values for each attribute, splitting them into unique text tokens and then computing the entropy across all values for an attribute.

Further security measures we implemented were to use \textit{HMAC-SHA-256} as a keyed hash function, apply random hashing, attribute salting, as well as balancing and randomly permuting Bloom filters after the masking step~\cite{niedermeyer_cryptanalysis_2014, schnell_randomized_2016}.
All these techniques mitigate basic dictionary and frequency attacks, as well as known cryptanalytic attacks on Bloom filters based on their Hamming weights at the time of writing.

\section{Results}

We used the readily available PADME PHT stations at the RWTH Aachen University, the University of Applied Sciences Mittweida and the University Medical Center Leipzig to validate our suggested approach to record linkage on distributed datasets.
We performed two experiments with two different datasets.
The source code for the generation of these datasets, as well as the software components we developed, is available from the authors.

For our first experiment, we used the North Carolina Voter Registration (NCVR) dataset\footnote{The data is available at \url{https://www.ncsbe.gov/results-data/voter-registration-data}. We used a snapshot of statewide voter registrations from February 16\textsuperscript{th}, 2023}.
This dataset has been used to test several record linkage applications in the past~\cite{kuzu_practical_2013, durham_composite_2014, durham_framework_2012, ranbaduge_securing_2020}.
Its popularity stems from the fact that it contains real-world data sourced from a large population.
It is a public record of registered voters in the state of North Carolina where a large number of personally identifiable attributes are tracked for every person.
These attributes include first, middle and last name, gender, year of birth, residential and mail address, party affiliation, phone number and supplemental information on the record itself.
Every person in the NCVR database has a unique identifier called NCID.
We performed data cleaning by only keeping records that abide by the following rules.

\begin{itemize}
\item person must have an active and verified record
\item person must have been at least \num{16} years old on their registration date
\item person must be no older than \num{120} years
\item record must not be marked as confidential
\item residential and mail zip codes must be either five or nine characters in length and must not be all zeros
\item phone numbers must not be all zeros
\end{itemize}

After this process, about \num{6.1} million records remain from the original \num{8.2} million records.
From the cleaned dataset, we selected first and last name, gender, year of birth and city of residence to be used for our record linkage experiment.

We created three files with approximately \num{5000} records each, where all files have \num{100} records in common.
To select records, we grouped all records by first and last name and counted the amount of records in each group, which we refer to as group size.
For each group size up to five, we then randomly selected records drawn from each group such that every group size contributes about a fifth of all records to the final dataset, which is composed of \num{14800} unique records across all files.
The distribution of records with respect to the group sizes they were drawn from is shown in Tab~\ref{tab:ncvd-draw}.

\begin{table}[!ht]
\centering
\caption{Amount of selected records in the final dataset with respect to group size based on distinct first and last names in the NCVR dataset}
\begin{tabular}{rrr}
\toprule
\textbf{Group size} & \textbf{\# Total records} & \textbf{\# Selected records} \\
\midrule
\num{1} & \num{2975253} & \num{2971} \\
\num{2} & \num{668560} & \num{2992} \\
\num{3} & \num{358608} & \num{2901} \\
\num{4} & \num{240640} & \num{2784} \\
\num{5} & \num{179465} & \num{2910} \\
\midrule
$\Sigma$ & \num{4422526} & \num{14558} \\
\bottomrule
\end{tabular}
\label{tab:ncvd-draw}
\end{table}

Finally, we selected \num{100} records at random and inserted them into each of the three files.
We inserted each of the remaining \num{14700} records into one of the three files at random.
The amount of records for each file is shown in Tab~\ref{tab:ncvd-files}.

\begin{table}[!ht]
\centering
\caption{Amount of records in each file for our record linkage experiment with the NCVR dataset}
\begin{tabular}{rr}
\toprule
\textbf{File} & \textbf{\# Records} \\
\midrule
\num{1} & \num{4981} \\
\num{2} & \num{4945} \\
\num{3} & \num{4832} \\
\bottomrule
\end{tabular}
\label{tab:ncvd-files}
\end{table}

We assigned the files to the participating stations at random, so that each station holds one of the three files.
The station administrators were provided a mapping to import the given files into the data schema of their on-premise E-PIX instance.
Knowing that only exact matches were supposed to be found, we set the match threshold for the train execution accordingly to \SI{100}{\percent}.
We used the NCIDs of predicted matching pairs to classify them as true or false positives.

We were able to identify all true matches while only accumulating three false positives, yielding a F1-score of \SI{99.5}{\percent}.
The false positives are a result of the limited subset of attributes we chose.
Upon review, we found that the affected records were sufficiently different from one another when observing all attributes of the NCVR dataset.
However, by chance alone, the false positives we identified were indeed identical by our choice of attributes.

For our second experiment, we generated a synthetic dataset with built-in typographic errors to validate the error-resistance of our approach.
We used German first names, surnames and city names with respective frequency information from publicly available data sources.
We used the GeCo framework~\cite{tran_geco_2013} to generate a file containing \num{1000} personal records.
The distribution of the aforementioned attribute values was modeled after their real-life frequencies.
We also added a randomly generated gender and birth date to each record.

Next, we generated two corrupted versions of the original file by randomly introducing typographic errors commonly found in the real world.
These include optical character recognition errors, phonetic errors, input errors, e.g. by pressing a neighboring key on a keyboard, and edit errors, such as random insertion, deletion or substitution of a character.
Finally, we shuffled the records between the three files in a way that ensured that matches could only be found between files, not in one and the same file.
As with the first dataset, we randomly assigned each file to one of the three participating stations and provided the mappings for the E-PIX import to all station administrators.

For our synthetic dataset, we opted for a more conservative matching threshold of \SI{70}{\percent}.
We found this threshold to be a reasonable choice in prior testing with similar datasets using our infrastructure.
In our train execution, we achieved an F1-score of \SI{99.3}{\percent}.
Out of \num{1157} possible true matches, we accumulated only four false positives and \num{12} false negatives, as represented in Tab~\ref{tab:result-synth}.
This result is impressive, considering the amount of true non-matches is bigger by about three orders of magnitude compared to the amount of matches.

\begin{table}[!ht]
\centering
\caption{
{Confusion matrix of predicted matches and non-matches at a \SI{70}{\percent} threshold}
}
\begin{tabular}{rrrr}
\toprule
 & True matches & True non-matches & $\Sigma$  \\
\midrule
Predicted matches & \num{1145} & \num{4} & \textbf{\num{1149}} \\
Predicted non-matches & \num{12} & \num{1142091} & \textbf{\num{1142103}} \\
\midrule
$\Sigma$ & \num{1157} & \num{1142095} &  \\
\bottomrule
\end{tabular}
\label{tab:result-synth}
\end{table}

\begin{figure}[!h]
\centering
\includegraphics[width=.5\textwidth]{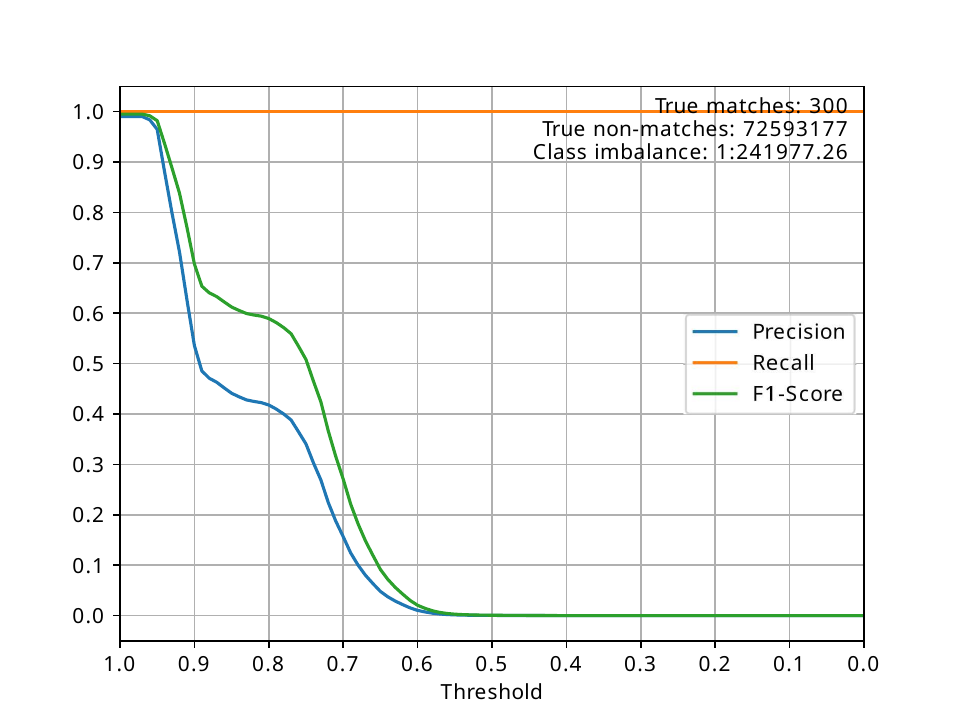}\hfill
\includegraphics[width=.5\textwidth]{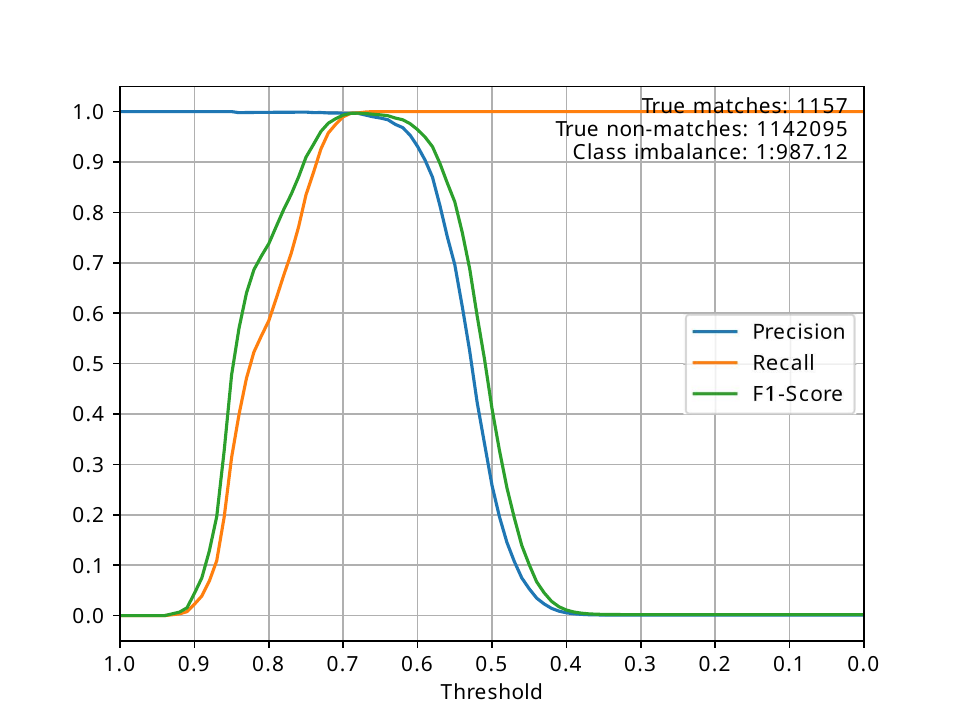}
\caption{Precision, recall and F1-score depending on the selected match threshold.
Measures were taken using our experimental datasets.
Samples were taken in \SI{1}{\percent} threshold increments.}
\label{fig:result-fscore}
\end{figure}

Finally, we conducted a statistical analysis on both datasets to evaluate our threshold choices.
We ran the same linkage procedures using a local setup with varying thresholds, ranging from \SI{0}{\percent} to \SI{100}{\percent} in \SI{1}{\percent} increments.
At every threshold, we computed precision, recall and F1-score.
The resulting plots are shown in Fig~\ref{fig:result-fscore}.

For the NCVR dataset, we observed a sharp decline in match quality around the \SI{95}{\percent} threshold.
The F1-score declined slowly between \SI{90}{\percent} and \SI{75}{\percent} before falling off drastically again and leveling off below \SI{60}{\percent}.
Since we chose records in groups based on distinct first and last names, a base level of similarity is present in a significant number of record pairs.
This explains the first drop-off as records with identical first and last names are being recognized as matches.
The second, significantly more prolonged drop-off is caused by record pairs that are decisively unlike one another.

For our synthetic dataset, we found that our threshold of choice at \SI{70}{\percent} was not far off the threshold which maximized the F1-score on our synthetic dataset.
At a threshold of \SI{68}{\percent}, the amount of false negatives dropped down to four, raising the F1-score to \SI{99.7}{\percent}.
Furthermore, we found that all true matches were classified as such at a threshold of \SI{66}{\percent}.
However, false positives increased drastically with lower thresholds.
On the opposite end, the first false positive was reported at a threshold of \SI{84}{\percent}, while about \SI{40}{\percent} of true matches were already being identified as such.

\section{Discussion}

\subsection{Security implications of the PHT}

The PHT concept allows the execution of arbitrary analysis tasks on medical data provided by healthcare institutions.
This is desirable for data scientists who act in good faith and wish to use the statistical tools and programming languages that they are familiar with, but this still poses a major threat to data protection and privacy.
Therefore PHT implementations must provide security guarantees to mitigate unauthorized access to and distribution of medical data.
We consider possible attack vectors for a malicious actor aiming to exfiltrate medical data using the PHT and show mitigations at the example of PADME and PHT-meDIC.

Both PADME and PHT-meDIC employ a manual approval process for incoming trains.
Station administrators must first inspect and confirm that they wish to execute a train at their station.
The code packaged inside a train is always public and should be used by station administrators to conclude what the train's purpose is and which data sources it aims to connect to.
This allows them to reject trains that have been sent to a station due to human error or malicious intent.

Furthermore, both PADME and PHT-meDIC require data sources to be defined manually.
For PADME, connection details must be provided before the train is executed.
This allows the data holders to configure their data sources in a way that mitigates illegal operations and to limit the train's view on the required data.
PHT-meDIC completely denies network access to the train.
Data sources are mounted into the train as a read-only volume.
This further mitigates the train's abilities to retrieve and share information outside of institutional borders.

Results of a train execution at a station must be securely stored, and PADME and PHT-meDIC choose different ways to ensure that no unauthorized party may be able to inspect results.
PADME employs the use of public-key cryptography.
The Central Service and all stations maintain their own keypairs which are used to encrypt trains and the results that are packaged into them after each train execution.
PHT-meDIC goes one step further by cryptographically signing trains to ensure that trains are only executed by trusted parties.
Results from a train execution at a station are not stored within the train but rather encrypted using homomorphic encryption.
Only the data scientist running the analysis task using PHT-meDIC may decrypt the results which ensures that analysis artifacts submitted to the central service cannot be read by its administrators.

At the time of writing, manual review and approval of train, manual configuration of data sources, restricting network access, result encryption and train signatures summarize the scope of security practices of both PADME and PHT-meDIC.
This serves to ensure that medical data is not leaked to untrusted parties, but puts a major responsibility on station administrators.
They must diligently inspect trains before executing them and ensure that the data sources are properly configured.
The public nature of both PHT implementations gives them the necessary tools to validate trains and approve or deny their execution on site.

\subsection{Attacks on the record linkage protocol}

Cryptanalysis of Bloom filters in privacy-preserving record linkage protocols has been conducted in the past with the consensus being that basic field-level Bloom filters provide no meaningful security for data that is meant to be obfuscated~\cite{niedermeyer_cryptanalysis_2014, kuzu_constraint_2011, christen_efficient_2017}.
We chose to adapt our use of Bloom filters for PPRL in line with current security recommendations, which make re-identification attacks infeasible.
Instead of providing a formal cryptanalysis of Bloom filters in our protocol, we present the scope of possible attack vectors that malicious actors could abuse in our approach.

There are three main actors who take part in a PPRL execution within the PHT.
The first actor is the data scientist.
Since they only interact with the Central Service, there is no way for them to get access to the underlying container infrastructure.
Their only entrypoint for attack is the PPRL container image, which can be freely adjusted.
This means that a malicious data scientist could include malicious code in the PPRL train image that is then executed at the participating stations.
However, within the context of the PPRL workflow as described in this paper, there is no way to obtain unauthorized access to confidential patient data.
The only information that could be leaked are the pseudonyms provided at every station, which inherently do not represent personally identifiable information.
Since the Resolver service is the sole component responsible for handling depseudonymized data, there is no feasible way of leaking confidential data from it using a malicious PPRL train image either.
Furthermore, the PADME PHT implementation allows station administrators to deny the execution of a train image that they do not trust.

The second actor is the station administrator.
Since they are restricted to executing or rejecting trains, there are no means for the station administrator to achieve unauthorized data access.
They have no access to the underlying container infrastructure that is managed by the station software and can therefore not leak results from executions at previous stations.
Since the PPRL train is not supposed to store sensitive data in the first place, it is not possible to infer patient data from other stations.

The third actor is the administrator of the CLU.
They are the weakest link, since we assume a trusted third party setting and therefore do not consider their entire attack scope in a practical setting.
However, a bad actor in this position has a few options.
The central linkage unit receives and processes bit vectors submitted by all stations.
We assume that a re-identification attack is too compute-intensive and therefore infeasible.
However, suppose there are matching bit vectors across all participating stations of a PPRL execution.
Given the metadata attached to these bit vectors, it would be an easy task for a malicious CLU administrator to make assumptions about the medical or treatment history of a patient that produced these bit vectors.
This greatly limits the search scope, though this knowledge is still useless without prior knowledge about personal patient details.
Furthermore, if the secret for the authenticated hash digest in the masking step is changed in every train execution, then inferring information during multiple train executions is not possible for the CLU administrator.

Another attack vector for a CLU administrator is to disrupt the PPRL execution.
Since the assignment of clients to a match session is handled by the central linkage unit, a malicious CLU administrator could assign submitted bit vectors to arbitrary match sessions, therefore impacting the usefulness of the reported results.
It is also possible to simply withhold match results or to inject false match results.

\subsection{Federated train execution}

Our proposed approach performs a train execution in two round trips.
This incremental workflow comes down to the fact that all participating stations are processed sequentially.
In theory, a federated workflow where all stations are processed simultaneously is also possible and would require a single round trip instead.

A train class for this federated approach is prepared.
All participating stations are selected and the train is dispatched.
Once at a station, the train performs the usual workflow of submitting pseudonyms to the Resolver service.
However, instead of terminating, the PPRL train queries the progress of the match session and waits for it to reach completion.
As soon as all stations have submitted their bit vectors and matching has been carried out, all trains submit a request to the Resolver to fetch the results from the broker.
The main challenge lies in coordinating all stations to perform their submission step in a similar time frame.
Suppose that in a setup with three stations, two execute the train immediately and one does so only an hour later after receiving, there is currently no way for the Broker service to tell when matching is to be considered finished.
In the worst case, the first two stations request their results too early, omitting potential matches with the third station.
A federated version of the train used in our approach is actively being worked on.

\subsection{Future improvements}

We used a simple measure for estimating attribute weight based on their discriminatory power.
Though weight estimation algorithms for record linkage exist~\cite{winkler2000using, blakely_probabilistic_2002}, none of them can be applied to blind record linkage, meaning that known algorithms which are mostly based on the Fellegi-Sunter model of record linkage cannot be applied to PPRL, since we do not have the datasets to be matched available to us.
In the future, we hope to develop and evaluate more advanced algorithms for blind attribute weight estimation.

The same applies to the threshold estimation.
We chose a threshold based on prior knowledge about the data to be matched and previous experiments.
A future area of research concerns the study of thresholds based on the chosen masking technique and input data.

While the GeCo framework is decent at generating authentic-looking datasets with a variety of options for common errors to induce, the data it generates is still synthetic.
We are actively working on cooperating with personal data providers to create use cases around our approach to PPRL to prove its effectiveness in real-world settings.

\section{Conclusion}

We presented an approach to perform record linkage on distributed datasets using the PHT.
The PHT enables data holding organizations to allow access to their personal records without leaking sensitive information beyond institutional borders.
Inferring information from masked records is infeasible with the security measures we chose to implement, keeping the chance of a successful re-identification attack at a minimum.
We validated this approach in two experiments with two different synthetic datasets with respect to the quality of the identified matches and the typographic error-resistance.
In the future, we aim to demonstrate our approach in real-life scenarios.

\printbibliography

@inproceedings{tran_geco_2013,
	address = {New York, NY, USA},
	series = {{CIKM} '13},
	title = {{GeCo}: an online personal data generator and corruptor},
	isbn = {978-1-4503-2263-8},
	shorttitle = {{GeCo}},
	url = {https://doi.org/10.1145/2505515.2508207},
	doi = {10.1145/2505515.2508207},
	urldate = {2022-07-18},
	booktitle = {Proceedings of the 22nd {ACM} international conference on {Information} \& {Knowledge} {Management}},
	publisher = {Association for Computing Machinery},
	author = {Tran, Khoi-Nguyen and Vatsalan, Dinusha and Christen, Peter},
	month = oct,
	year = {2013},
	pages = {2473--2476},
}

@article{vatsalan_evaluation_2014,
	title = {An {Evaluation} {Framework} for {Privacy}-{Preserving} {Record} {Linkage}},
	volume = {6},
	copyright = {Copyright (c) 2014 Dinusha Vatsalan, Peter Christen, Christine M. O'Keefe, Vassilios S. Verykios},
	issn = {2575-8527},
	url = {https://journalprivacyconfidentiality.org/index.php/jpc/article/view/636},
	doi = {10.29012/jpc.v6i1.636},
	language = {en},
	number = {1},
	urldate = {2022-02-14},
	journal = {Journal of Privacy and Confidentiality},
	author = {Vatsalan, Dinusha and Christen, Peter and O'Keefe, Christine M. and Verykios, Vassilios S.},
	month = jun,
	year = {2014},
	note = {Number: 1},
}

@misc{niedermeyer_cryptanalysis_2014,
	address = {Rochester, NY},
	title = {Cryptanalysis of {Basic} {Bloom} {Filters} {Used} for {Privacy} {Preserving} {Record} {Linkage}},
	url = {https://papers.ssrn.com/abstract=3530867},
	doi = {10.2139/ssrn.3530867},
	language = {en},
	urldate = {2022-07-11},
	author = {Niedermeyer, Frank and Steinmetzer, Simone and Kroll, Martin and Schnell, Rainer},
	month = jun,
	year = {2014},
}

@inproceedings{schnell_randomized_2016,
	title = {Randomized {Response} and {Balanced} {Bloom} {Filters} for {Privacy} {Preserving} {Record} {Linkage}},
	doi = {10.1109/ICDMW.2016.0038},
	booktitle = {2016 {IEEE} 16th {International} {Conference} on {Data} {Mining} {Workshops} ({ICDMW})},
	author = {Schnell, Rainer and Borgs, Christian},
	month = dec,
	year = {2016},
	note = {ISSN: 2375-9259},
	pages = {218--224},
}

@article{yigzaw_secure_2017,
	title = {Secure and scalable deduplication of horizontally partitioned health data for privacy-preserving distributed statistical computation},
	volume = {17},
	copyright = {2016 The Author(s).},
	issn = {1472-6947},
	url = {https://bmcmedinformdecismak.biomedcentral.com/articles/10.1186/s12911-016-0389-x},
	doi = {10.1186/s12911-016-0389-x},
	language = {en},
	number = {1},
	urldate = {2022-06-08},
	journal = {BMC Medical Informatics and Decision Making},
	author = {Yigzaw, Kassaye Yitbarek and Michalas, Antonis and Bellika, Johan Gustav},
	month = dec,
	year = {2017},
	note = {Number: 1
Publisher: BioMed Central},
	pages = {1--19},
}

@article{nguyen_privacy-preserving_2020,
	title = {Privacy-{Preserving} {Record} {Linkage} of {Deidentified} {Records} {Within} a {Public} {Health} {Surveillance} {System}: {Evaluation} {Study}},
	volume = {22},
	shorttitle = {Privacy-{Preserving} {Record} {Linkage} of {Deidentified} {Records} {Within} a {Public} {Health} {Surveillance} {System}},
	url = {https://www.jmir.org/2020/6/e16757},
	doi = {10.2196/16757},
	language = {EN},
	number = {6},
	urldate = {2022-09-22},
	journal = {Journal of Medical Internet Research},
	author = {Nguyen, Long and Stoové, Mark and Boyle, Douglas and Callander, Denton and McManus, Hamish and Asselin, Jason and Guy, Rebecca and Donovan, Basil and Hellard, Margaret and El-Hayek, Carol},
	month = jun,
	year = {2020},
	note = {Company: Journal of Medical Internet Research
Distributor: Journal of Medical Internet Research
Institution: Journal of Medical Internet Research
Label: Journal of Medical Internet Research
Publisher: JMIR Publications Inc., Toronto, Canada},
	pages = {e16757},
}

@article{randall_privacy-preserving_2014,
	series = {Special {Issue} on {Informatics} {Methods} in {Medical} {Privacy}},
	title = {Privacy-preserving record linkage on large real world datasets},
	volume = {50},
	issn = {1532-0464},
	url = {https://www.sciencedirect.com/science/article/pii/S1532046413001949},
	doi = {10.1016/j.jbi.2013.12.003},
	language = {en},
	urldate = {2021-12-14},
	journal = {Journal of Biomedical Informatics},
	author = {Randall, Sean M. and Ferrante, Anna M. and Boyd, James H. and Bauer, Jacqueline K. and Semmens, James B.},
	month = aug,
	year = {2014},
	pages = {205--212},
}

@article{nobrega_blockchain-based_2021,
	title = {Blockchain-based {Privacy}-{Preserving} {Record} {Linkage}: enhancing data privacy in an untrusted environment},
	volume = {102},
	issn = {0306-4379},
	shorttitle = {Blockchain-based {Privacy}-{Preserving} {Record} {Linkage}},
	url = {https://www.sciencedirect.com/science/article/pii/S0306437921000661},
	doi = {10.1016/j.is.2021.101826},
	language = {en},
	urldate = {2022-10-06},
	journal = {Information Systems},
	author = {Nóbrega, Thiago and Pires, Carlos Eduardo S. and Nascimento, Dimas Cassimiro},
	month = dec,
	year = {2021},
	pages = {101826},
}

@article{christen_critique_2022,
	title = {A critique and attack on “{Blockchain}-based privacy-preserving record linkage”},
	volume = {108},
	issn = {0306-4379},
	url = {https://www.sciencedirect.com/science/article/pii/S0306437921001320},
	doi = {10.1016/j.is.2021.101930},
	language = {en},
	urldate = {2022-10-06},
	journal = {Information Systems},
	author = {Christen, Peter and Schnell, Rainer and Ranbaduge, Thilina and Vidanage, Anushka},
	month = sep,
	year = {2022},
	pages = {101930},
}

@article{nobrega_explanation_2022,
	title = {Explanation and answers to critiques on: {Blockchain}-based {Privacy}-{Preserving} {Record} {Linkage}},
	volume = {108},
	issn = {0306-4379},
	shorttitle = {Explanation and answers to critiques on},
	url = {https://www.sciencedirect.com/science/article/pii/S0306437921001356},
	doi = {10.1016/j.is.2021.101935},
	language = {en},
	urldate = {2022-10-06},
	journal = {Information Systems},
	author = {Nóbrega, Thiago and Pires, Carlos Eduardo S. and Nascimento, Dimas Cassimiro},
	month = sep,
	year = {2022},
	pages = {101935},
}

@article{stammler_mainzelliste_2020,
	title = {Mainzelliste {SecureEpiLinker} ({MainSEL}): {Privacy}-{Preserving} {Record} {Linkage} using {Secure} {Multi}-{Party} {Computation}},
	issn = {1367-4803},
	shorttitle = {Mainzelliste {SecureEpiLinker} ({MainSEL})},
	url = {https://doi.org/10.1093/bioinformatics/btaa764},
	doi = {10.1093/bioinformatics/btaa764},
	urldate = {2021-12-10},
	journal = {Bioinformatics},
	author = {Stammler, Sebastian and Kussel, Tobias and Schoppmann, Phillipp and Stampe, Florian and Tremper, Galina and Katzenbeisser, Stefan and Hamacher, Kay and Lablans, Martin},
	month = sep,
	year = {2020},
	pages = {btaa764},
}

@article{contiero_epilink_2005,
	title = {The {EpiLink} {Record} {Linkage} {Software}},
	volume = {44},
	copyright = {Schattauer GmbH},
	issn = {0026-1270, 2511-705X},
	url = {http://www.thieme-connect.de/DOI/DOI?10.1055/s-0038-1633924},
	doi = {10.1055/s-0038-1633924},
	language = {en},
	number = {1},
	urldate = {2021-12-16},
	journal = {Methods of Information in Medicine},
	author = {Contiero, P. and Tittarelli, A. and Tagliabue, G. and Maghini, A. and Fabiano, S. and Crosignani, P. and Tessandori, R.},
	year = {2005},
	note = {Publisher: Schattauer GmbH},
	pages = {66--71},
}

@article{welten_privacy-preserving_2022,
	title = {A {Privacy}-{Preserving} {Distributed} {Analytics} {Platform} for {Health} {Care} {Data}},
	volume = {61},
	copyright = {Georg Thieme Verlag KG Rüdigerstraße 14, 70469 Stuttgart, Germany},
	issn = {0026-1270, 2511-705X},
	url = {http://www.thieme-connect.de/DOI/DOI?10.1055/s-0041-1740564},
	doi = {10.1055/s-0041-1740564},
	language = {en},
	number = {S 01},
	urldate = {2022-09-22},
	journal = {Methods of Information in Medicine},
	author = {Welten, Sascha and Mou, Yongli and Neumann, Laurenz and Jaberansary, Mehrshad and Ucer, Yeliz Yediel and Kirsten, Toralf and Decker, Stefan and Beyan, Oya},
	month = jun,
	year = {2022},
	note = {Publisher: Georg Thieme Verlag KG},
	pages = {e1--e11},
}

@article{bialke_mosaic_2015,
	title = {{MOSAIC} – {A} {Modular} {Approach} to {Data} {Management} in {Epidemiological} {Studies}},
	volume = {54},
	copyright = {Georg Thieme Verlag KG Stuttgart · New York},
	issn = {0026-1270, 2511-705X},
	url = {http://www.thieme-connect.de/DOI/DOI?10.3414/ME14-01-0133},
	doi = {10.3414/ME14-01-0133},
	language = {en},
	number = {04},
	urldate = {2021-12-13},
	journal = {Methods of Information in Medicine},
	author = {Bialke, M. and Bahls, T. and Havemann, C. and Piegsa, J. and Weitmann, K. and Wegner, T. and Hoffmann, W.},
	month = jul,
	year = {2015},
	note = {Publisher: Georg Thieme Verlag KG},
	pages = {364--371},
}

@article{bialke_workflow-driven_2015,
	title = {A workflow-driven approach to integrate generic software modules in a {Trusted} {Third} {Party}},
	volume = {13},
	copyright = {2015 Bialke et al.},
	issn = {1479-5876},
	url = {https://translational-medicine.biomedcentral.com/articles/10.1186/s12967-015-0545-6},
	doi = {10.1186/s12967-015-0545-6},
	language = {en},
	number = {1},
	urldate = {2021-12-13},
	journal = {Journal of Translational Medicine},
	author = {Bialke, Martin and Penndorf, Peter and Wegner, Tim and Bahls, Thomas and Havemann, Christoph and Piegsa, Jens and Hoffmann, Wolfgang},
	month = dec,
	year = {2015},
	note = {Number: 1
Publisher: BioMed Central},
	pages = {1--8},
}

@inproceedings{kuzu_constraint_2011,
	address = {Berlin, Heidelberg},
	series = {Lecture {Notes} in {Computer} {Science}},
	title = {A {Constraint} {Satisfaction} {Cryptanalysis} of {Bloom} {Filters} in {Private} {Record} {Linkage}},
	isbn = {978-3-642-22263-4},
	doi = {10.1007/978-3-642-22263-4_13},
	language = {en},
	booktitle = {Privacy {Enhancing} {Technologies}},
	publisher = {Springer},
	author = {Kuzu, Mehmet and Kantarcioglu, Murat and Durham, Elizabeth and Malin, Bradley},
	editor = {Fischer-Hübner, Simone and Hopper, Nicholas},
	year = {2011},
	pages = {226--245},
}

@inproceedings{christen_efficient_2017,
	address = {Cham},
	series = {Lecture {Notes} in {Computer} {Science}},
	title = {Efficient {Cryptanalysis} of {Bloom} {Filters} for {Privacy}-{Preserving} {Record} {Linkage}},
	isbn = {978-3-319-57454-7},
	doi = {10.1007/978-3-319-57454-7_49},
	language = {en},
	booktitle = {Advances in {Knowledge} {Discovery} and {Data} {Mining}},
	publisher = {Springer International Publishing},
	author = {Christen, Peter and Schnell, Rainer and Vatsalan, Dinusha and Ranbaduge, Thilina},
	editor = {Kim, Jinho and Shim, Kyuseok and Cao, Longbing and Lee, Jae-Gil and Lin, Xuemin and Moon, Yang-Sae},
	year = {2017},
	pages = {628--640},
}

@techreport{schnell_novel_2011,
	address = {Rochester, NY},
	type = {{SSRN} {Scholarly} {Paper}},
	title = {A {Novel} {Error}-{Tolerant} {Anonymous} {Linking} {Code}},
	url = {https://papers.ssrn.com/abstract=3549247},
	language = {en},
	number = {ID 3549247},
	urldate = {2022-03-14},
	institution = {Social Science Research Network},
	author = {Schnell, Rainer and Bachteler, Tobias and Reiher, Jörg},
	month = nov,
	year = {2011},
	doi = {10.2139/ssrn.3549247},
}

@article{beyan_distributed_2020,
	title = {Distributed {Analytics} on {Sensitive} {Medical} {Data}: {The} {Personal} {Health} {Train}},
	volume = {2},
	issn = {2641-435X},
	shorttitle = {Distributed {Analytics} on {Sensitive} {Medical} {Data}},
	url = {https://direct.mit.edu/dint/article/2/1-2/96-107/9997},
	doi = {10.1162/dint\_a\_00032},
	language = {en},
	number = {1-2},
	urldate = {2021-10-27},
	journal = {Data Intelligence},
	author = {Beyan, Oya and Choudhury, Ananya and van Soest, Johan and Kohlbacher, Oliver and Zimmermann, Lukas and Stenzhorn, Holger and Karim, Md. Rezaul and Dumontier, Michel and Decker, Stefan and da Silva Santos, Luiz Olavo Bonino and Dekker, Andre},
	month = jan,
	year = {2020},
	pages = {96--107},
}

@article{moncada-torres_vantage6_2021,
	title = {{VANTAGE6}: an open source {priVAcy} {preserviNg} {federaTed} {leArninG} {infrastructurE} for {Secure} {Insight} {eXchange}},
	volume = {2020},
	issn = {1942-597X},
	shorttitle = {{VANTAGE6}},
	url = {https://www.ncbi.nlm.nih.gov/pmc/articles/PMC8075508/},
	urldate = {2022-03-08},
	journal = {AMIA Annual Symposium Proceedings},
	author = {Moncada-Torres, Arturo and Martin, Frank and Sieswerda, Melle and Van Soest, Johan and Geleijnse, Gijs},
	month = jan,
	year = {2021},
	pmid = {33936462},
	pmcid = {PMC8075508},
	pages = {870--877},
}

@article{durham_composite_2014,
	title = {Composite {Bloom} {Filters} for {Secure} {Record} {Linkage}},
	volume = {26},
	issn = {1558-2191},
	doi = {10.1109/TKDE.2013.91},
	number = {12},
	journal = {IEEE Transactions on Knowledge and Data Engineering},
	author = {Durham, Elizabeth A. and Kantarcioglu, Murat and Xue, Yuan and Toth, Csaba and Kuzu, Mehmet and Malin, Bradley},
	month = dec,
	year = {2014},
	note = {Conference Name: IEEE Transactions on Knowledge and Data Engineering},
	pages = {2956--2968},
}

@article{mou_distributed_2021,
	title = {Distributed {Skin} {Lesion} {Analysis} {Across} {Decentralised} {Data} {Sources}},
	url = {https://ebooks.iospress.nl/doi/10.3233/SHTI210179},
	doi = {10.3233/SHTI210179},
	urldate = {2022-09-22},
	journal = {Public Health and Informatics},
	author = {Mou, Yongli and Welten, Sascha and Jaberansary, Mehrshad and Ucer Yediel, Yeliz and Kirsten, Toralf and Decker, Stefan and Beyan, Oya},
	year = {2021},
	note = {Publisher: IOS Press},
	pages = {352--356},
}

@article{welten_multi-institutional_2022,
	title = {Multi-{Institutional} {Breast} {Cancer} {Detection} {Using} a {Secure} {On}-{Boarding} {Service} for {Distributed} {Analytics}},
	volume = {12},
	copyright = {http://creativecommons.org/licenses/by/3.0/},
	issn = {2076-3417},
	url = {https://www.mdpi.com/2076-3417/12/9/4336},
	doi = {10.3390/app12094336},
	language = {en},
	number = {9},
	urldate = {2023-02-08},
	journal = {Applied Sciences},
	author = {Welten, Sascha and Hempel, Lars and Abedi, Masoud and Mou, Yongli and Jaberansary, Mehrshad and Neumann, Laurenz and Weber, Sven and Tahar, Kais and Ucer Yediel, Yeliz and Löbe, Matthias and Decker, Stefan and Beyan, Oya and Kirsten, Toralf},
	month = jan,
	year = {2022},
	note = {Number: 9
Publisher: Multidisciplinary Digital Publishing Institute},
	pages = {4336},
}

@article{welten_synthetic_2022,
	title = {Synthetic rainfall data generator development through decentralised model training},
	volume = {612},
	issn = {0022-1694},
	url = {https://www.sciencedirect.com/science/article/pii/S0022169422007818},
	doi = {10.1016/j.jhydrol.2022.128210},
	language = {en},
	urldate = {2023-02-08},
	journal = {Journal of Hydrology},
	author = {Welten, Sascha and Holt, Adrian and Hofmann, Julian and Schelter, Lennart and Klopries, Elena-Maria and Wintgens, Thomas and Decker, Stefan},
	month = sep,
	year = {2022},
	pages = {128210},
}

@book{winkler2000using,
  title={Using the {EM} algorithm for weight computation in the {Fellegi-Sunter} model of record linkage},
  author={Winkler, William E},
  year={2000},
  publisher={US Bureau of the Census Washington, DC},
  pages={12}
}

@book{voigt_eu_2017,
	address = {Cham},
	title = {The {EU} {General} {Data} {Protection} {Regulation} ({GDPR})},
	isbn = {978-3-319-57959-7},
	url = {http://link.springer.com/10.1007/978-3-319-57959-7},
	language = {en},
	urldate = {2023-02-13},
	publisher = {Springer International Publishing},
	author = {Voigt, Paul and von dem Bussche, Axel},
	year = {2017},
	doi = {10.1007/978-3-319-57959-7},
}

@incollection{winkler_chapter_2009,
	series = {Handbook of {Statistics}},
	title = {Chapter 14 - {Record} {Linkage}},
	volume = {29},
	url = {https://www.sciencedirect.com/science/article/pii/S016971610800014X},
	language = {en},
	urldate = {2023-02-14},
	booktitle = {Handbook of {Statistics}},
	publisher = {Elsevier},
	author = {Winkler, William E.},
	editor = {Rao, C. R.},
	month = jan,
	year = {2009},
	doi = {10.1016/S0169-7161(08)00014-X},
	pages = {351--380},
}

@article{goecks_galaxy_2010,
	title = {Galaxy: a comprehensive approach for supporting accessible, reproducible, and transparent computational research in the life sciences},
	volume = {11},
	issn = {1474-760X},
	shorttitle = {Galaxy},
	url = {https://doi.org/10.1186/gb-2010-11-8-r86},
	doi = {10.1186/gb-2010-11-8-r86},
	language = {en},
	number = {8},
	urldate = {2023-02-21},
	journal = {Genome Biology},
	author = {Goecks, Jeremy and Nekrutenko, Anton and Taylor, James and {The Galaxy Team}},
	month = aug,
	year = {2010},
	pages = {R86},
}

@article{crosswell_elixir_2012,
	title = {{ELIXIR}: a distributed infrastructure for {European} biological data},
	volume = {30},
	issn = {0167-7799, 1879-3096},
	shorttitle = {{ELIXIR}},
	url = {https://www.cell.com/trends/biotechnology/abstract/S0167-7799(12)00017-0},
	doi = {10.1016/j.tibtech.2012.02.002},
	language = {English},
	number = {5},
	urldate = {2023-02-27},
	journal = {Trends in Biotechnology},
	author = {Crosswell, Lindsey C. and Thornton, Janet M.},
	month = may,
	year = {2012},
	pmid = {22417641},
	note = {Publisher: Elsevier},
	pages = {241--242},
}

@article{gaye_datashield_2014,
	title = {{DataSHIELD}: taking the analysis to the data, not the data to the analysis},
	volume = {43},
	issn = {0300-5771},
	shorttitle = {{DataSHIELD}},
	url = {https://doi.org/10.1093/ije/dyu188},
	doi = {10.1093/ije/dyu188},
	number = {6},
	urldate = {2022-03-08},
	journal = {International Journal of Epidemiology},
	author = {Gaye, Amadou and Marcon, Yannick and Isaeva, Julia and LaFlamme, Philippe and Turner, Andrew and Jones, Elinor M and Minion, Joel and Boyd, Andrew W and Newby, Christopher J and Nuotio, Marja-Liisa and Wilson, Rebecca and Butters, Oliver and Murtagh, Barnaby and Demir, Ipek and Doiron, Dany and Giepmans, Lisette and Wallace, Susan E and Budin-Ljøsne, Isabelle and Oliver Schmidt, Carsten and Boffetta, Paolo and Boniol, Mathieu and Bota, Maria and Carter, Kim W and deKlerk, Nick and Dibben, Chris and Francis, Richard W and Hiekkalinna, Tero and Hveem, Kristian and Kvaløy, Kirsti and Millar, Sean and Perry, Ivan J and Peters, Annette and Phillips, Catherine M and Popham, Frank and Raab, Gillian and Reischl, Eva and Sheehan, Nuala and Waldenberger, Melanie and Perola, Markus and van den Heuvel, Edwin and Macleod, John and Knoppers, Bartha M and Stolk, Ronald P and Fortier, Isabel and Harris, Jennifer R and Woffenbuttel, Bruce HR and Murtagh, Madeleine J and Ferretti, Vincent and Burton, Paul R},
	month = dec,
	year = {2014},
	pages = {1929--1944},
}

@incollection{christen_data_2012,
	address = {Berlin, Heidelberg},
	series = {Data-{Centric} {Systems} and {Applications}},
	title = {The {Data} {Matching} {Process}},
	isbn = {978-3-642-31164-2},
	url = {https://doi.org/10.1007/978-3-642-31164-2_2},
	language = {en},
	urldate = {2021-12-17},
	booktitle = {Data {Matching}: {Concepts} and {Techniques} for {Record} {Linkage}, {Entity} {Resolution}, and {Duplicate} {Detection}},
	publisher = {Springer},
	author = {Christen, Peter},
	editor = {Christen, Peter},
	year = {2012},
	doi = {10.1007/978-3-642-31164-2_2},
	pages = {23--35},
}

@article{blakely_probabilistic_2002,
	title = {Probabilistic record linkage and a method to calculate the positive predictive value},
	volume = {31},
	issn = {0300-5771},
	url = {https://doi.org/10.1093/ije/31.6.1246},
	doi = {10.1093/ije/31.6.1246},
	number = {6},
	urldate = {2023-02-27},
	journal = {International Journal of Epidemiology},
	author = {Blakely, Tony and Salmond, Clare},
	month = dec,
	year = {2002},
	pages = {1246--1252},
}

@article{kuzu_practical_2013,
	title = {A practical approach to achieve private medical record linkage in light of public resources},
	volume = {20},
	issn = {1067-5027},
	url = {https://doi.org/10.1136/amiajnl-2012-000917},
	doi = {10.1136/amiajnl-2012-000917},
	number = {2},
	urldate = {2023-03-06},
	journal = {Journal of the American Medical Informatics Association},
	author = {Kuzu, Mehmet and Kantarcioglu, Murat and Durham, Elizabeth Ashley and Toth, Csaba and Malin, Bradley},
	month = mar,
	year = {2013},
	pages = {285--292},
}

@mastersthesis{durham_framework_2012,
	title = {A framework for accurate, efficient private record linkage},
	url = {https://ir.vanderbilt.edu/handle/1803/11417},
	language = {en},
	urldate = {2022-02-15},
	author = {Durham, Elizabeth Ashley},
	month = apr,
	year = {2012},
	note = {Accepted: 2020-08-22T00:02:52Z},
	type = {Doctoral Thesis},
	school = {Vanderbilt University}
}

@inproceedings{ranbaduge_securing_2020,
	address = {New York, NY, USA},
	series = {{CIKM} '20},
	title = {Securing {Bloom} {Filters} for {Privacy}-preserving {Record} {Linkage}},
	isbn = {978-1-4503-6859-9},
	url = {https://doi.org/10.1145/3340531.3412105},
	doi = {10.1145/3340531.3412105},
	urldate = {2022-03-01},
	booktitle = {Proceedings of the 29th {ACM} {International} {Conference} on {Information} \& {Knowledge} {Management}},
	publisher = {Association for Computing Machinery},
	author = {Ranbaduge, Thilina and Schnell, Rainer},
	month = oct,
	year = {2020},
	pages = {2185--2188},
}

@misc{herr_bringing_2022,
	title = {Bringing the {Algorithms} to the {Data} -- {Secure} {Distributed} {Medical} {Analytics} using the {Personal} {Health} {Train} ({PHT}-{meDIC})},
	url = {http://arxiv.org/abs/2212.03481},
	doi = {10.48550/arXiv.2212.03481},
	urldate = {2023-09-11},
	publisher = {arXiv},
	author = {Herr, Marius de Arruda Botelho and Graf, Michael and Placzek, Peter and König, Florian and Bötte, Felix and Stickel, Tyra and Hieber, David and Zimmermann, Lukas and Slupina, Michael and Mohr, Christopher and Biergans, Stephanie and Akgün, Mete and Pfeifer, Nico and Kohlbacher, Oliver},
	month = dec,
	year = {2022},
	note = {arXiv:2212.03481 [cs]},
}

\end{document}